\documentstyle[prc,aps,preprint]{revtex}
\begin{document} \draft
\date{\today}
\title{The coupling constants g$_{\rho\pi\gamma}$ and g$_{\omega\pi\gamma}$
as derived from QCD sum rules}

\author{A. Gokalp~\thanks{agokalp@metu.edu.tr} and
        O. Yilmaz~\thanks{oyilmaz@metu.edu.tr}}
\address{ {\it Physics Department, Middle East Technical University,
06531 Ankara, Turkey}}
\maketitle

\begin{abstract}
We employ QCD sum rules to calculate the coupling constants
g$_{\rho\pi\gamma}$ and g$_{\omega\pi\gamma}$ by studying the
three point ${\rho\pi\gamma}$- and ${\omega\pi\gamma}$-correlation
functions. Our results for the decay widths
$\Gamma(\rho^0\rightarrow \pi^0\gamma)$ and
$\Gamma(\omega\rightarrow \pi^0\gamma)$ calculated using the
obtained coupling constants are in good agreement with the
experimental values of these decay widths.
\end{abstract}

\thispagestyle{empty} ~~~~\\ 
\newpage
\setcounter{page}{1}

Radiative transitions of the type $V\rightarrow P\gamma$ where V
and P belong the lowest multiplets of vector (V) and pseudoscalar
(P) mesons have been a subject of continuous interest both
theoretically and experimentally \cite{R1}. These transitions have
been considered from the point of view of a large variety of
theoretical models, such as phenomenological quark models
\cite{R2}, potential models \cite{R3}, bag models \cite{R4}, and
effective Lagrangian approaches \cite{R5}. All these approaches
provide effective methods of investigation of these hadronic
phenomena for which a formulation for the application of QCD from
the first principles has not been possible  so far. The effective
Lagrangian approach provide a framework to study in general the
physics of light neutral vector mesons, $\rho^0$, $\omega$ and
$\phi$, by combining the Vector Meson Dominance and Chiral
Dynamics which are the two principles governing low energy QCD in
a suitably constructed effective Lagrangian \cite{R6}.

On the other hand, vector meson-pseudoscalar meson-photon
$VP\gamma$-vertex also plays a role in photoproduction reactions
of vector mesons on nucleons. Although, at sufficiently high
energies and low momentum transfers electromagnetic production of
vector mesons on nucleon targets has been explained by Pomeron
exchange models, at low energies near threshold scalar and
pseudoscalar meson exchange mechanisms become important \cite{R7}.
For the photoproduction reactions involving $\rho^0$ and $\omega$
mesons, the effective coupling constants g$_{\rho\pi\gamma}$ and
g$_{\omega\pi\gamma}$ are among the physical inputs that are used
in the analyzes of these reactions. In these studies, an effective
Lagrangian describing the $VP\gamma$-vertex is assumed, which also
defines the coupling constant g$_{VP\gamma}$, and these coupling
constants are then determined utilizing the experimental decay
widths $\Gamma(V\rightarrow P\gamma)$ of the vector mesons.
However, it should be noted that in these decays the four-momentum
of the pseudoscalar meson P is time-like, ${p^\prime}^2>0$,
whereas in the pseudoscalar exchange amplitude contributing to the
photoproduction of vector mesons it is space-like
${p^\prime}^2<0$. Therefore, it is of interest to study the
effective coupling constants g$_{VP\gamma}$ from another point of
view as well.

In this work, we estimate the coupling constants
g$_{\rho\pi\gamma}$ and g$_{\omega\pi\gamma}$ by employing QCD sum
rules which provide an efficient and model-independent method to
study many hadronic observables, such as decay constants and form
factors, in terms of nonperturbative contributions proportional to
the quark and gluon condensates \cite{R8,R9}. Using the techniques
of QCD sum rules, the nonperturbative QCD physics is incorporated
systematically as power corrections in the short-distance operator
product expansion.

In order to derive the QCD sum rule for the coupling constant
g$_{V\pi\gamma}$ where V denotes $\rho^0$ or $\omega$ meson, we
begin by considering the three point correlation function
\begin{equation}\label{e1}
  T_{\mu\nu}(p,p^\prime)=\int d^{4}x d^4y e^{ip^\prime\cdot y}e^{-ip\cdot x}
  <o|T\{j_\mu^\gamma(0)j_{\nu}^V(x)j_{5}(y)\}|0>
\end{equation}
where the interpolating currents $j_{\nu}^V$ for$\rho^0$ and
$\omega$ meson are
$j_{\nu}^{\rho}=\frac{1}{2}(\overline{u}\gamma_{\nu}u-\overline{d}\gamma_{\nu}d)$,
$j_{\nu}^{\omega}=\frac{1}{6}(\overline{u}\gamma_{\nu}u+\overline{d}\gamma_{\nu}d)$,
respectively,
$j_{5}=\frac{1}{2}(\overline{u}i\gamma_{5}u-\overline{d}i\gamma_{5}d)$
is the interpolating current for $\pi^0$ \cite{R8}, and
$j_{\mu}^{\gamma}=e_u\overline{u}\gamma_{\mu}u+e_d\overline{d}\gamma_{\mu}d$,
where $e_u$ and $e_d$ denote the quark charges, is the quark
electromagnetic current. In accordance with QCD sum rules
techniques, we consider the three point correlation function
$T_{\mu\nu}(p,p^\prime)$ in the Euclidean region defined by
$p^2=-Q^2\sim -1~~GeV^2$, ${p^\prime}^2=-{Q^\prime}^2\sim
-1~~GeV^2$.

The theoretical part of the sum rule for the coupling constant
g$_{V\pi\gamma}$ is obtained in terms of QCD degrees of freedom by
calculating the perturbative contribution and the power
corrections from operators of different dimensions to the three
point correlation function. In the region $Q^2,~{Q^\prime}^2\sim
1~~GeV^2$ the perturbative contribution can be approximated by the
lowest order free-quark loop diagram shown in Fig. 1-a.
Furthermore, we consider the power corrections from operators of
different dimensions, resulting in contributions to the three
point correlation function that are proportional to the terms
$<\overline{q}q>$, $<\sigma\cdot G>$ and $<(\overline{q}q)^2>$. We
do not consider the gluon condensate contribution proportional to
$<G^2>$ since it is estimated to be negligible for light quark
systems. The calculations of the power corrections are performed
in the fixed point gauge \cite{R10}. We work in the SU(2) flavour
context with m$_u$=m$_d$=m$_q$, moreover we perform our
calculations of the perturbative and power correction
contributions in the limit m$_q=0$. In this limit, the
perturbative bare-loop diagram does not make any contribution, and
only operators of dimensions d=3 and d=5 make contributions that
are proportional to $<\overline{q}q>$ and $<\sigma\cdot G>$,
respectively. The relevant Feynman diagrams for the calculation of
these power corrections are shown in Fig. 1-b and Fig. 1-c.

We then turn to the calculation of the three point correlation
function through phenomenological considerations. The vertex
function $T_{\mu\nu}(p,p^\prime)$ satisfies a double dispersion
relation. In general such a dispersion relation can be written in
three ways by choosing two of the three channels. For our purpose,
we choose the vector and pseudoscalar channels and by saturating
this dispersion relation by the lowest lying meson states in these
channels we obtain the physical part of the sum rule as
 \begin{equation}\label{e2}
  T_{\mu\nu}(p,p^\prime)=\frac{<0|j_{\nu}^V|V>
  <V(p)|j_\mu^\gamma|\pi(p^\prime)><\pi|j_{5}|0>}
  {(p^2-m^2_V)({p^\prime}^2-m^2_\pi)}+...
\end{equation}
where the contributions from the higher states and the continuum
is denoted by dots. In this expression, the overlap amplitudes for
vector and pseudoscalar mesons are $<0|j_{\nu}^V|V>=\lambda_V u_V$
where $u_V$ is the polarization vector of the vector meson and
$<\pi|j_{5}|0>=\lambda_\pi$. The matrix element of the
electromagnetic current is given as
\begin{equation}\label{e8}
<V(p)|j_\mu^\gamma|\pi(p^\prime)>=
-i\frac{e}{m_V}g_{V\pi\gamma}K(q^2)\varepsilon^{\mu\alpha\beta\delta}p_\alpha
u_\beta q_\delta
\end{equation}
where $q=p-p^\prime$ and K(0)=1. This expression defines the
coupling constant g$_{V\pi\gamma}$ through the effective
Lagrangian
\begin{equation}\label{e9}
{\cal
L}^{eff.}_{V\pi\gamma}=\frac{e}{m_V}g_{V\pi\gamma}\varepsilon^{\mu\nu\alpha\beta}
\partial_\mu V_\nu\partial_\alpha A_\beta\pi^0
\end{equation}
describing the $V\pi\gamma$-vertex \cite{R7}.

After performing the double Borel transform with respect to the
variables $Q^2$ and ${Q^\prime}^2$, we obtain the sum rule for the
coupling constant  g$_{V\pi\gamma}$
\begin{eqnarray}\label{e10}
g_{V\pi\gamma}=\frac{3m_V}{\lambda_V\lambda_\pi}
  e^{\frac{m_V^2}{M^2}}e^{\frac{m_\pi^2}{{M^\prime}^2}}C_V<\overline{q}q>
    \left (-\frac{3}{4}+\frac{5}{32}m_0^2\frac{1}{M^2}
    -\frac{3}{32}m_0^2\frac{1}{{M^\prime}^2}\right )
\end{eqnarray}
where we use the relation $<\sigma\cdot G>=m_0^2<\overline{q}q>$.
The constant $C_V$ that results in our calculation is $C_V=1$ for
$\rho^0$ meson and $C_V=3$ for $\omega$ meson. For the numerical
evaluation of the sum rule we use the values $m_0^2=0.8~~GeV^2$,
$<\overline{u}u>=-0.014~~GeV^3$ \cite{R11}, and
$m_\rho=0.770~~GeV$, $m_\omega=0.782~~GeV$, $m_\pi=0.138~~GeV$
\cite{R12} . For the overlap amplitude for the vector meson
states, we use the values that are obtained from the experimental
leptonic decay widths  \cite{R12} by noting that neglecting the
electron mass the  $e^+ e^-$ decay width of vector meson is given
as $\Gamma (V\rightarrow e^+
e^-)=\frac{4\pi\alpha^2}{3}\frac{\lambda_V^2}{m_V^3}$, this way we
obtain the values $\lambda_\rho=0.118~~GeV^2$ and
$\lambda_\omega=0.036~~GeV^2$. We note that these values do obey
the SU(3) relation $\lambda_\rho=3\lambda_\omega$ within 10$\%$
accuracy. The overlap amplitude $\lambda_\pi$ for the $\pi$ meson
state is given by the relation
$\lambda_{\pi}=f_{\pi}\frac{m_\pi^2}{m_u+m_d}$ \cite{R14}. We use
the experimental value $f_\pi$=0.132 GeV and the physical mass
$m_{\pi^0}$=0.138 GeV along with $m_u+m_d$=0.014 GeV, and obtain
this amplitude as $\lambda_\pi$=0.18 GeV$^2$. In order to analyze
the dependence of the coupling constant g$_{V\pi\gamma}$ on the
Borel parameters $M^2$ and ${M^\prime}^2$, we study independent
variations of $M^2$ and ${M^\prime}^2$ in the interval
$0.6~~GeV^2\leq M^2,{M^\prime}^2\leq 1.4~~GeV^2$ as these limits
determine the allowed interval for the vector channel \cite{R13}.
The variation of the coupling constant g$_{\rho\pi\gamma}$ and
g$_{\omega\pi\gamma}$ as a function of Borel parameters $M^2$ for
different values of ${M^\prime}^2$ is shown in Fig. 2 and in Fig.
3, respectively. The examination of these figures indicate that
the sum rule is quite stable with these reasonable variations of
$M^2$ and ${M^\prime}^2$. Besides those due to variations of $M^2$
and ${M^\prime}^2$, the other sources contributing to the
uncertainty in the coupling constants are the uncertainties in the
estimated values of the vacuum condensates. If we take these
uncertainties into account by a conservative estimate, we obtain
the coupling constants g$_{\rho\pi\gamma}=0.63\pm 0.07$ and
g$_{\omega\pi\gamma}=1.85\pm 0.15$. These values of the coupling
constant are consistent with their values used in the analysis of
$\rho^0$ and $\omega$ photoproduction reactions through
pseudoscalar exchange amplitudes which are
g$_{\rho\pi\gamma}=0.54$ and g$_{\omega\pi\gamma}=1.82$,
respectively  \cite{R15}. If we use the effective Lagrangian given
in Eq. 4, then the decay width for $V\rightarrow\pi^0\gamma$ is
obtained as
\begin{equation}\label{e6}
  \Gamma(V\rightarrow\pi^0\gamma)=\frac{\alpha}{24}
  \frac{(m_V^2-m_{\pi^0}^2)^3}{m_V^5}g_{V\pi\gamma}^2 ~~.
\end{equation}
Therefore, from our analysis we determine
$\Gamma(V\rightarrow\pi^0\gamma)$ decay widths for $\rho^0$ and
$\omega$ mesons as  $\Gamma(\rho^0\rightarrow\pi^0\gamma)=(84\pm
15)$ KeV and $\Gamma(\omega\rightarrow\pi^0\gamma)=(740\pm 80)$
KeV. The measured decay widths \cite{R12} are
$\Gamma(\rho^0\rightarrow\pi^0\gamma)=(102\pm 26)$ KeV and
$\Gamma(\omega\rightarrow\pi^0\gamma)=(717\pm 43)$ KeV which
roughly follow the SU(3) prediction for their ratio. Our results
are in good agreement with the experimental values of these decay
widths. We also note that the electromagnetic decays $V\rightarrow
P\gamma$ of vector mesons in the flavour SU(3) sector was studied
previously \cite{R14} by employing the method of QCD sum rules in
the presence of the external electromagnetic field. Our results
which are obtained by QCD sum rules utilizing three point
correlation functions are consistent with the values obtained in
that analysis and therefore supplements the study of these decays
using QCD sum rules method.

\begin{center}
{\bf ACKNOWLEDGMENTS}
\end{center}

We thank  T. M. Aliev for helpful discussions during the course of
our work.


\newpage

{\bf Figure Captions:}

\begin{description}

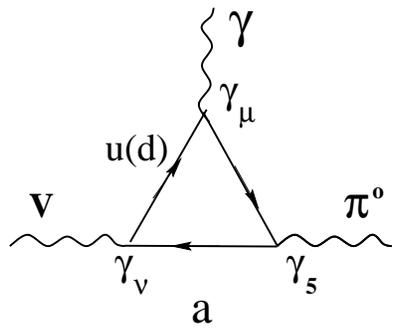
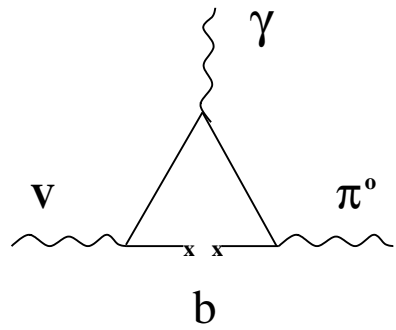
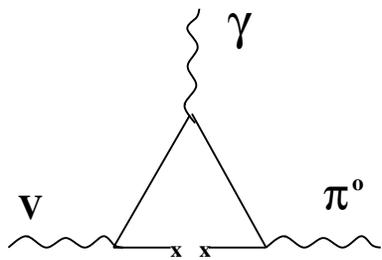
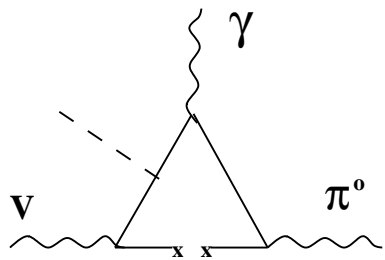
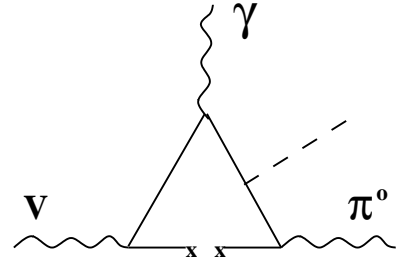
\item[{\bf Figure 1}:] Feynman Diagrams for the
$V\pi\gamma$-vertex: a- bare loop diagram, b- d=3 operator
corrections, c- d=5 operator corrections. The dotted lines denote
gluons.

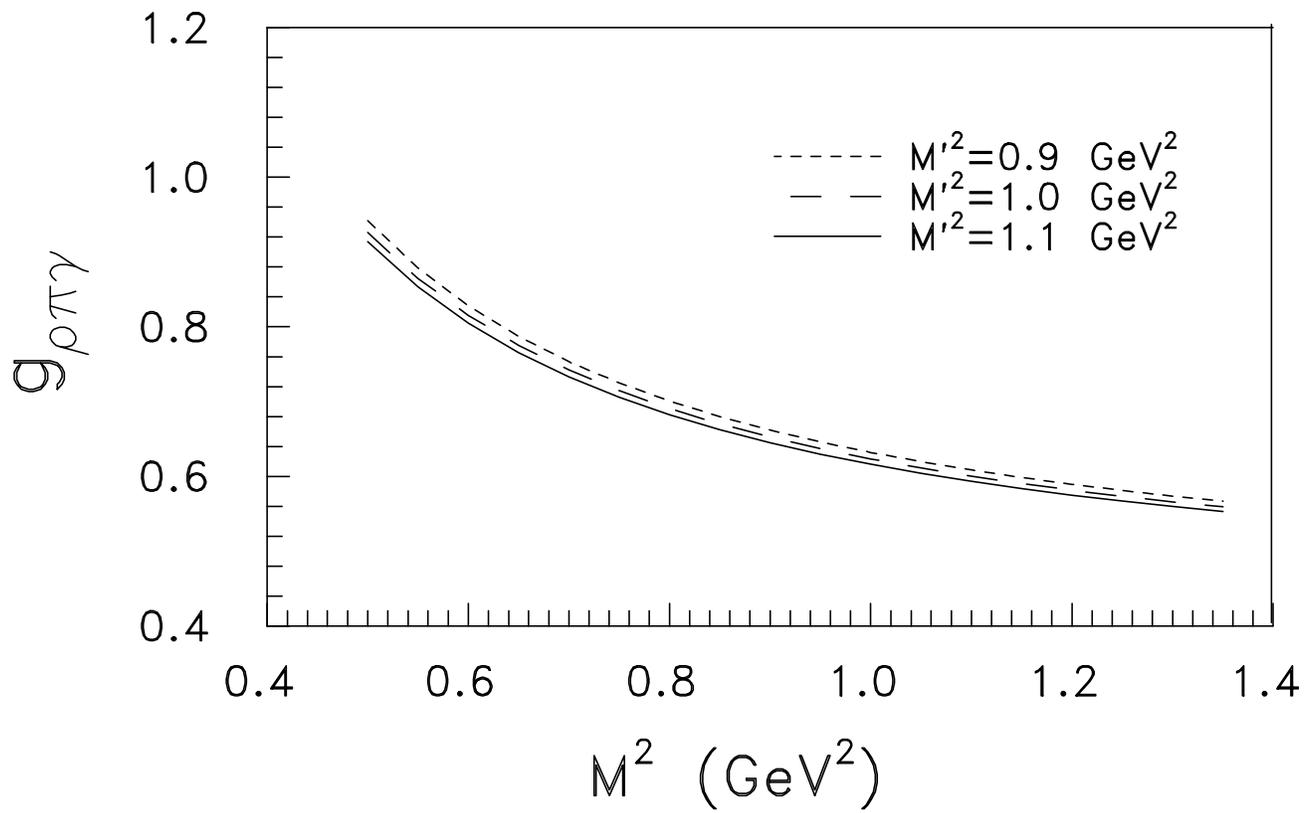
\item[{\bf Figure 2}:] The coupling constant $g_{\rho\pi\gamma}$
as a function of the Borel parameter $M^2$ for different values of
${M^\prime}^2$.

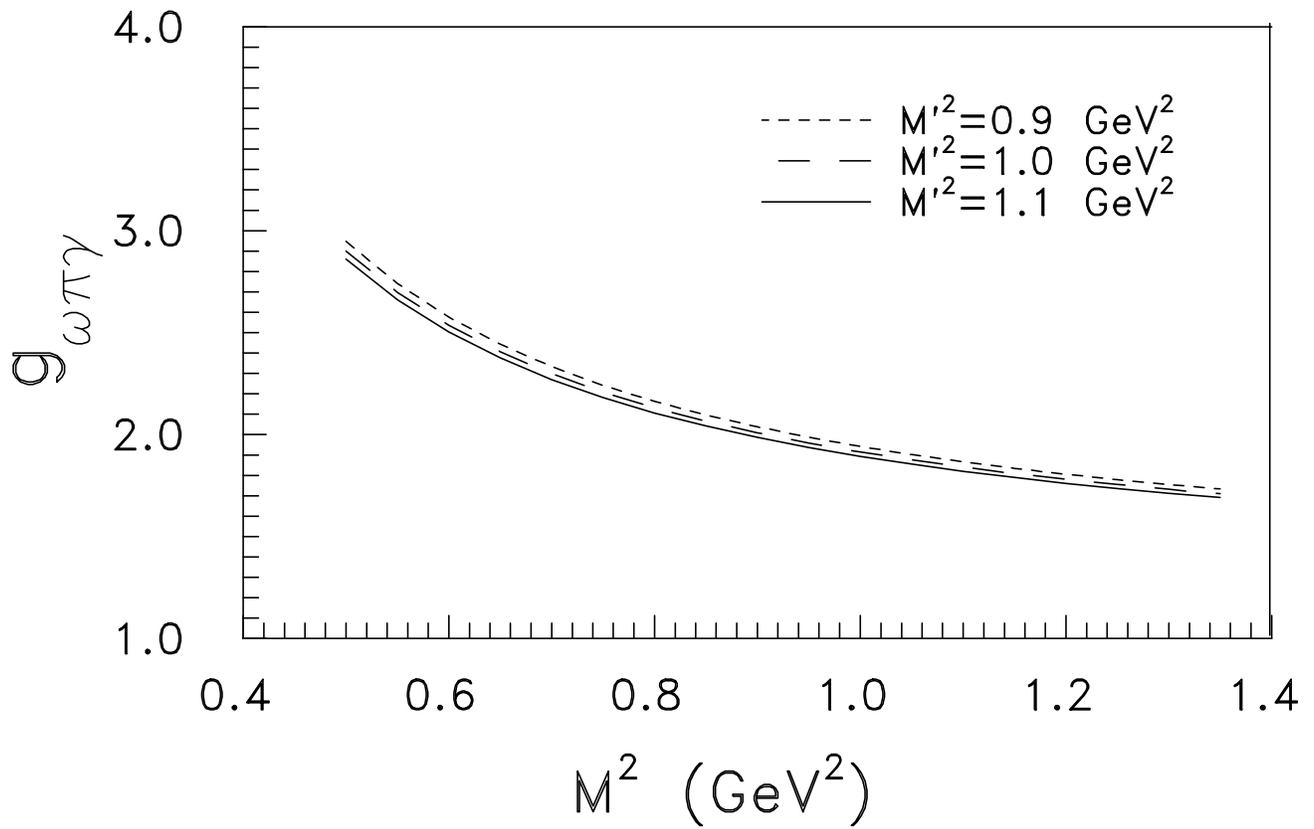
\item[{\bf Figure 3}:] The coupling constant $g_{\omega\pi\gamma}$
as a function of the Borel parameter $M^2$ for different values of
${M^\prime}^2$.
\end{description}

\end{document}